# Approximating CCCV charging using SOC-dependent tapered charging power constraints in long-term microgrid planning

Hassan Zahid Butt
*Student Member, IEEE*
Department of Electrical and Computer Engineering
University of Houston
Houston, TX, USA
hbutt@uh.edu

Xingpeng Li
*Senior Member, IEEE*
Department of Electrical and Computer Engineering
University of Houston
Houston, TX, USA
xli82@uh.edu

*Abstract*— Traditional long-term microgrid planning models assume constant power charging for battery energy storage systems (BESS), overlooking efficiency losses that occur toward the end of charge due to rising internal resistance. While this issue can be mitigated at the cell level using constant current–constant voltage (CCCV) charging, it is impractical at the pack level in large-scale systems. However, battery management systems and inverter controls can emulate this effect by tapering charging power at high state-of-charge (SOC) levels, trading off charging speed for improved efficiency and reduced thermal stress. Ignoring this behavior in planning models can lead to undersized batteries and potential reliability issues. This paper proposes a tractable and scalable approach to approximate CCCV behavior using SOC-dependent tapered charging power (TCP) constraints. A MATLAB-based proof of concept demonstrates the energy delivery and efficiency benefits of tapering. The method is integrated into a long-term planning framework and evaluated under a synthetic load and solar profile. Results show tapering significantly affects BESS sizing, cost, and reliability under dynamic operating conditions that demand fast charging. These findings highlight tapering as a critical modeling factor for accurately capturing BESS performance in long-term microgrid planning.

NOMENCLATURE

Sets:
| | |
|---|---|
| $Y$ | Set of total number of years |
| $D$ | Set of representative days in a single year |
| $T$ | Set of hourly time periods in a single day |
| $B$ | Set of SOC bands |

Indices:
| | |
|---|---|
| $y$ | Year $y$, an element of set $Y$ |
| $d$ | Day $d$, an element of set $D$ |
| $t$ | Time period $t$, an element of set $T$ |
| $b$ | SOC band Index $b$, an element of set $B$ |

Parameters:
| | |
|---|---|
| $P_{y,d,t}^{load}$ | Total load (MW) in year $y$, day $d$, & hour $t$ |
| $P_{y,d,t}^{PV}$ | Solar power capacity factor in year $y$, day $d$, & hour $t$ |
| $C_{PV}^{capital}$ | PV capital cost factor ($/MW) |
| $C_{BESS}^{capital}$ | BESS capital cost factor ($/MWh) |
| $\gamma_{PV}^{rep}$ | PV replacement cost as a percent of capital cost |
| $T_{BESS}^{chg}$ | Duration of BESS charging (h) |
| $T_{BESS}^{dchg}$ | Duration of BESS discharging (h) |
| $\eta_{chg}$ | BESS charging energy efficiency |
| $\eta_{dchg}$ | BESS discharging energy efficiency |
| $\eta_{PV}^{init}$ | Initial PV conversion efficiency |
| $SOC_{max}$ | Maximum state of charge limit for BESS |
| $SOC_{min}$ | Minimum state of charge limit for BESS |
| $\delta_{PV}^{deg}$ | PV efficiency degradation rate per annum |
| $C_{LS}^{penalty}$ | Penalty cost factor for load shedding ($/MW) |
| $M_{SOC}$ | Big-M constant used to activate/relax SOC bands |
| $M_{BESS}$ | Big-M constant used to activate/relax BESS operational constraints |
| $Y_{MG}$ | Total microgrid planning years |
| $\alpha$ | Scaling factor for repeating load and solar profiles |
| $\tau_b^{upper}$ | Upper threshold for SOC band |
| $\tau_b^{lower}$ | Lower threshold for SOC band |
| $\beta_b$ | Tapering Factor for SOC band |

Variables:
| | |
|---|---|
| $S_{PV}$ | PV system size (MW) |
| $S_{BESS}$ | BESS energy capacity (MWh) |
| $P_{y,d,t}^{pv\_curt}$ | PV power curtailed (MW) in year $y$, day $d$, & hour $t$ |
| $P_{y,d,t}^{LS}$ | Load shed (MW) in year $y$, day $d$, & hour $t$ |
| $P_{y,d,t}^{chg}$ | BESS charge power (MW) in year $y$, day $d$, & hour $t$ |
| $P_{y,d,t}^{dchg}$ | BESS discharge power (MW) in year $y$, day $d$, & hour $t$ |
| $E_{BESS}^{init}$ | Initial BESS energy level (MWh) |
| $E_{y,d,t}^{BESS}$ | Energy level of BESS (MWh) in year $y$, day $d$, & hour $t$ |
| $C_{PV}^{deg}$ | PV degradation cost ($) |
| $U_{y,d,t}^{chg}$ | BESS charging status in year $y$, day $d$, & hour $t$ |
| $U_{y,d,t}^{dchg}$ | BESS discharging status in year $y$, day $d$, & hour $t$ |
| $SOC_{y,d,t}$ | BESS state of charge at year $y$, day $d$, & hour $t$ |
| $u_{b,y,d,t}^{soc\_band}$ | SOC band selection binary indicator |
| $\eta_y^{PV}$ | PV conversion efficiency in year $y$ |

## I. INTRODUCTION

Microgrids (MGs) are increasingly central to the modern energy landscape, enabling decentralized, resilient power systems that align with global sustainability targets [1]. By supporting localized generation and storage, MGs reduce reliance on centralized infrastructure, minimize transmission losses, and improve energy access, particularly in remote or underserved areas [2].

The integration of renewable energy sources, especially solar photovoltaics (PV), has driven MG adoption due to their low operating costs, scalability, and environmental benefits [3]-[4]. However, the inherent intermittency of solar generation necessitates reliable energy storage to ensure supply continuity, particularly for off-grid systems [5]. To address this intermittency challenge, lithium-ion battery energy storage systems (BESS) are widely used due to their high energy density, flexible deployment, and rapid response capabilities, offering advantages over alternatives such as pumped hydro storage [6]-[8]. When paired with PV, BESS enhances energy autonomy, improves system resilience, and supports decarbonization goals [9].

Despite these advantages, high capital costs remain a significant barrier to widespread PV-plus-BESS deployment. Oversizing the system increases investment cost, while undersizing compromises reliability [10]. Most long-term and short-term microgrid optimization planning models simplify battery charging behavior by assuming a constant power charging profile [11]-[12]. While this assumption reduces model complexity, it overlooks key physical and operational characteristics of real-world batteries. At the cell level, batteries are typically charged using constant current–constant voltage (CCCV) protocols [13]-[14]. This two-phase approach i.e., high current charging followed by a tapering phase near full charge, helps reduce I²R losses, minimizes thermal stress, and increases charging efficiency [15]. Although implementing CCCV at the system level is challenging due to the complexity of managing large battery packs, modern battery management systems (BMS) and inverter controls can approximate this tapering behavior by reducing charging power as the state of charge (SOC) increases using state-of-the-art communication protocols [16]. Fig. 1 shows a multi-stage constant current charging strategy, where the charging current decreases stepwise with increasing SOC [17].

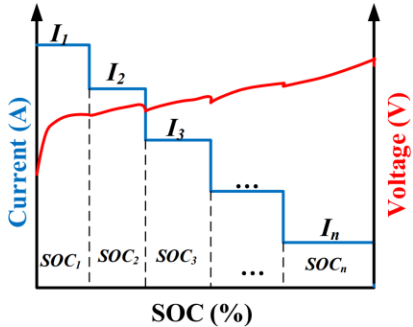

Fig. 1. An example of a multi-stage constant current charging waveforms with SOC-based transition [17].

Neglecting this tapering effect in long-term planning may result in overestimated charging flexibility, leading to undersized BESS investments and potential reliability issues. Moreover, tapering introduces a critical trade-off: improved efficiency at the cost of slower charging. Capturing this trade-off is essential for making informed investment decisions. Moreover, as the use of renewable energy resources increase, the behavior of end-user profiles becomes more dynamic. This introduces greater variability in load profiles and presents significant challenges for system planning and operation [18].

The rest of this paper is organized as follows. Section II presents a literature review, highlighting existing gaps and the contributions made by this study. Section III describes the methodology, including a proof-of-concept simulation, the baseline optimization framework, and the SOC-dependent tapered charging power (TCP) constraints. Section IV discusses case studies and key results. Finally, Section V concludes the paper and outlines potential directions for future research.

## II. LITERATURE REVIEW

BESSs have become central to microgrid design and planning, with extensive research focused on optimal sizing, techno-economic evaluation, and degradation-aware investment strategies. For example, Gholami et al. [19] proposed reliability-driven BESS sizing approaches, while Shen et al. [20] explored expansion planning incorporating centralized and distributed storage. Other works have modeled BESS replacement and degradation costs using linear or SOC-based metrics [21]-[22].

However, a common limitation in these long-term planning models is the oversimplification of battery charging behavior. Most studies assume constant power charging throughout the SOC range, which fails to reflect physical characteristics such as internal resistance rise at high SOC and resulting energy losses [23]. This can lead to overestimated flexibility and underestimated BESS capacity needs.

At the control level, CCCV charging protocols are well-established and recognized for reducing thermal stress and prolonging battery life. Several modeling studies [17], [24]-[25] have simulated CCCV effects on Li-ion cells, showing efficiency degradation and thermal impacts during high SOC charging. Yet, these insights are rarely embedded in optimization models for long-term microgrid planning, where charging power limits are typically static or unconstrained.

More recently, researchers have begun incorporating C-rate dependent degradation [26] and efficiency-aware constraints [27] in short-term operational models, particularly for frequency regulation or EV charging contexts. These models recognize that charging behavior affects not only energy losses but also the long-term health of the BESS. Still, their scope remains limited to dispatch horizons, with little attention to investment-stage planning.

To the best of our knowledge, no existing work has integrated SOC-dependent tapered charging constraints, which mimic the practical tapering behavior enforced by BMS in a long-term microgrid planning framework. This paper addresses that gap by introducing a tractable approximation of CCCV charging through SOC-dependent power limits and assessing its implications on BESS sizing, objective cost, and system reliability.

### A. Research Gaps & Contributions

Despite significant progress in long-term microgrid planning, most planning frameworks assume constant power charging of BESS, which overestimates its charging flexibility and fails to account for energy losses and thermal impacts that occur at high SOC due to increased internal resistance. Furthermore, its influence on investment-stage decisions, such as optimal BESS sizing, PV utilization, and system reliability, remains underexplored.

This paper addresses these gaps by introducing a tractable and scalable method to approximate CCCV charging behavior using SOC-dependent TCP constraints. These constraints dynamically limit charging power beyond specific SOC thresholds, emulating the tapering phase observed in real-world BMS-controlled systems. The contributions of this work are summarized as follows:
i. A SOC-dependent tapering mechanism is proposed to approximate CCCV charging by applying piecewise power limits, enabling scalable integration into long-term planning models.
ii. A MATLAB-based simulation is used to validate tapering benefits by quantifying efficiency gains and energy losses reduction under varying internal resistance.
iii. A multi-year optimization framework is developed with embedded tapering constraints to co-optimize PV and BESS sizing under charging flexibility limits.

### III. METHODOLOGY

Tapered charging, commonly implemented through CCCV protocols at the cell level, plays a vital role in enhancing battery efficiency, longevity, and safety. As internal resistance rises at high SOC, tapering the charging current helps reduce I²R losses, minimize heat generation, and slow thermal degradation. It also improves charging efficiency by reducing energy losses, mitigates C-rate-induced aging, and supports better cell balancing through more stable voltage profiles. While these benefits are well-established at the operational level, tapering introduces a trade-off by extending charging duration, which may impact performance in applications requiring fast turnaround. To evaluate this trade-off in the context of long-term planning, we first simulate and quantify the operational effects of tapering under controlled conditions and then embed the tapering mechanism into the planning optimization model.

#### A. Proof-of-Concept via MATLAB Simulation

To understand the practical benefits of tapering during the charging phase, a MATLAB/Simulink-based proof-of-concept model was developed as shown in Fig. 2.

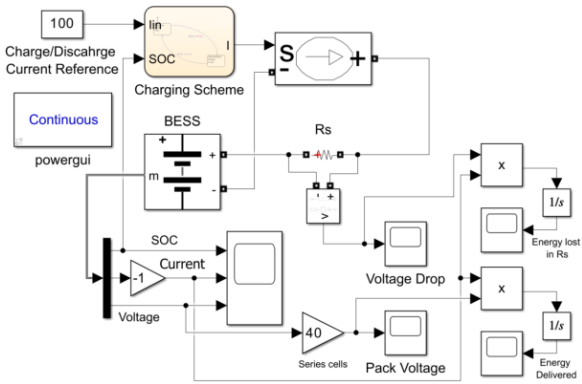

Fig. 2: Simulink model for analyzing the impact of charging strategies and internal resistance on energy loss and delivery.

The system simulates a BESS that is charged and discharged at a 1C rate under two charging strategies: conventional constant current charging, and a tapered charging profile that reduces current as SOC increases. The BESS model incorporates an internal resistance component (denoted Rs) to account for ohmic losses. The BESS selected is a 100Ah Li-Ion pack with 40 cells connected in series. The charging scheme is toggled between a constant 1C charging, and a tapered charging logic that begins at 1C and tapers through discrete steps (0.5C, 0.2C, 0.1C) as the SOC crosses 80%, 90%, and 95% respectively, and switches to discharging once the BESS is fully charged. Further, two distinct values of Rs (0.001Ω and 0.01Ω) are tested to evaluate the sensitivity of BESS performance to internal resistance.

Equations (1) - (6) define the key performance metrics used to evaluate charging strategies. Equation (1) computes the total energy delivered to the battery pack by integrating the product of pack voltage and current over time. Equation (2) calculates energy lost as heat across the internal resistance Rs. The net usable energy, given by (3), is the difference between delivered and lost energy. Charging efficiency, defined in (4), expresses the ratio of net energy to total delivered energy. To quantify the benefit of tapering, (5) measures the reduction in energy loss compared to constant current charging, while (6) calculates the corresponding improvement in charging efficiency. Together, these metrics offer a comprehensive view of energy flow and losses under different charging profiles.

$$E_{delivered} = \int V_{pack} * I_{pack} \quad (1)$$
$$E_{loss} = \int V_R * I_{pack} \quad (2)$$
$$E_{net} = E_{delivered} - E_{loss} \quad (3)$$
$$\eta_{chg} = \frac{E_{net}}{E_{delivered}} * 100\% \quad (4)$$
$$\Delta E_{loss_{kWh}} = E_{loss} - E_{loss,taper} \quad (5)$$
$$\eta_{gain} = \eta_{taper} - \eta_{no_{taper}} \quad (6)$$

Simulation results for all five test cases are presented in Table I. Case 1 represents an ideal battery with no internal resistance. Cases 2 and 3 simulate a battery with Rs = 0.001 Ω under constant current and tapered charging, respectively. Cases 4 and 5 repeat these two schemes with Rs = 0.01 Ω. Results indicate that the benefits of tapering become more significant as internal resistance increases. At a low resistance of Rs = 0.001 Ω, tapering (Case 3) reduced energy loss by only 5 kWh compared to constant current charging (Case 2), with a marginal efficiency gain of 0.01%. However, at a higher resistance of Rs = 0.01 Ω, tapering (Case 5) reduced energy loss by 49 kWh compared to its constant current counterpart (Case 4), resulting in an efficiency improvement of 0.06%.

TABLE I
MATLAB SIMULATION RESULTS FOR DIFFERENT CHARGING SCHEMES

| Case # | Charging time | Energy delivered kWh | Energy loss kWh | Net Energy kWh | $\eta_{chg}$ |
|---|---|---|---|---|---|
| 1 | 3600s | 47,500 | 0 | 47,500 | 100% |
| 2 | 3600s | 48,940 | 36 | 48,904 | 99.93% |
| 3 | 6263s | 48,470 | 31 | 47,220 | 99.94% |
| 4 | 3600s | 61,900 | 360 | 61,540 | 99.42% |
| 5 | 6263s | 59,680 | 311 | 59,369 | 99.48% |

Although the absolute efficiency gain may seem small, these benefits become significant over repeated cycles in daily operations in their lifespan. The results suggest that tapering is particularly valuable in systems with high internal resistance, such as second-life batteries, where managing resistive losses is critical for energy efficiency and thermal management. This validates the necessity of the inclusion of tapering constraints in long-term microgrid planning models.

## B. Baseline Optimization Framework (without Tapering)

The baseline model minimizes the total microgrid cost over a multi-year planning horizon, encompassing capital investment, degradation penalties, and reliability considerations. The objective function is formulated as:

$$\min \left(S_{PV}C_{PV}^{capital} + S_{BESS}C_{BESS}^{capital}\right) + C_{PV}^{deg}Y_{MG} + \alpha \sum_{y \in Y}\sum_{d \in D}\sum_{t \in T}(P_{y,d,t}^{LS}C_{LS}^{penalty}) \quad (7)$$

This formulation ensures investment decisions are optimized while discouraging load shedding through high penalty costs. The constraints governing this model are as follows:

$$P_{y,d,t}^{dchg} + (\eta_y^{PV}P_{y,d,t}^{PV}S_{pv}) + P_{y,d,t}^{LS} = P_{y,d,t}^{load} + P_{y,d,t}^{chg} + P_{y,d,t}^{pv_{curt}} \quad (8)$$

$$0 \leq P_{y,d,t}^{pv_{curt}} \leq \eta_y^{PV}P_{y,d,t}^{PV}S_{PV} \quad (9)$$

$$0 \leq P_{y,d,t}^{LS} \leq P_{y,d,t}^{load} \quad (10)$$

$$C_{PV}^{deg} = \gamma_{PV}^{rep} \cdot \left(C_{PV}^{capital} \cdot S_{PV} \cdot \delta_{PV}^{deg}\right) \quad (11)$$

$$\eta_y^{PV} = \begin{cases} \eta_{init}^{PV}, & y = 1 \\ \eta_{y-1}^{PV}(1 - \delta_{PV}^{deg}), & y > 1 \end{cases} \quad (12)$$

$$SOC_{min}S_{BESS} \leq E_{y,d,t}^{BESS} \leq SOC_{max}S_{BESS} \quad (13)$$

$$SOC_{min}S_{BESS} \leq E_{BESS}^{init} \leq SOC_{max}S_{BESS} \quad (14)$$

$$U_{y,d,t}^{chg} + U_{y,d,t}^{dchg} \leq 1 \quad (15)$$

$$0 \leq P_{y,d,t}^{chg} \leq U_{y,d,t}^{chg}\frac{S_{BESS}}{T_{BESS}^{chg}} \quad (16)$$

$$0 \leq P_{y,d,t}^{dchg} \leq U_{y,d,t}^{dchg}\frac{S_{BESS}}{T_{BESS}^{dchg}} \quad (17)$$

$$E_{y,d,t}^{BESS} = \begin{cases} E_{BESS}^{init} + \left(\eta_{chg}P_{y,d,1}^{chg} - \frac{P_{y,d,1}^{dchg}}{\eta_{dchg}}\right), & t = 1 \\ E_{y,d,t-1}^{BESS} + \left(\eta_{chg}P_{y,d,t}^{chg} - \frac{P_{y,d,t}^{dchg}}{\eta_{dchg}}\right), & t > 1 \end{cases} \quad (18)$$

The power balance equation (8) ensures that the total energy demand is met by the combination of PV generation and BESS discharging, while also accounting for load shedding, PV curtailment, and charging power drawn by the BESS. Equation (9), which governs PV curtailment, calculates any surplus generation, ensuring that excess solar energy beyond the system's immediate consumption is either stored or curtailed, thereby avoiding infeasible overgeneration. Likewise, (10) handles load shedding by accurately capturing any unmet demand, thus maintaining model feasibility.

To support long-term system viability, degradation mechanisms are incorporated. The cost associated with PV degradation is defined in (11), where the financial loss is tied to system capacity and assumes a fixed annual drop in efficiency. Equation (12) models the decline in PV performance by applying a yearly degradation rate across the planning horizon. BESS operational constraints are enforced in (13) and (14), which confine the energy level within the prescribed SOC boundaries. Realistic battery behavior is further ensured in (15), which restricts concurrent charging and discharging, while (16) and (17) cap the power levels during these operations to prevent degradation from excessive charge rates. Finally, energy tracking for the battery is handled in (18), which updates the BESS energy state dynamically based on the efficiency-adjusted charge and discharge values.

## C. Proposed SOC-Dependent Tapering Constraints

To approximate CCCV charging behavior in a scalable and tractable manner, we introduce SOC-dependent TCP constraints that dynamically limit BESS charging power based on the battery's SOC. These constraints reduce the allowable charging rate in predefined SOC bands, mimicking the tapering phase of CCCV protocols without requiring nonlinear voltage-current relationships.

The total SOC range can be divided into discrete bands (e.g., 0–80%, 80–90%, 90–95%, 95–100%), each associated with a maximum charging power limit. Binary variables are used to track the active SOC band, and the corresponding tapering factor is applied accordingly. The formulation is as follows:

$$SOC_{y,d,t} = \frac{e_{y,d,t}^{bess}}{s_{bess}} \quad (19)$$

$$\sum_{b \in B} u_{b,y,d,t}^{soc_{band}} = U_{y,d,t}^{chg} \quad (20)$$

$$\begin{cases} SOC_{y,d,t} \geq \tau_b^{lower} - \left(1 - u_{b,y,d,t}^{soc_{band}}\right)M_{SOC} \\ SOC_{y,d,t} \leq \tau_b^{upper} + \left(1 - u_{b,y,d,y}^{soc_{band}}\right)M_{SOC} \end{cases} \quad (21)$$

$$\begin{cases} P_{y,d,t}^{chg} \leq U_{y,d,t}^{chg} M_{BESS} \\ P_{y,d,t}^{chg} \leq \sum_{b \in B} u_{b,y,d,y}^{soc_{band}} \beta_b \left(\frac{s_{bess}}{T_{BESS}^{chg}}\right) \end{cases} \quad (22)$$

The SOC at any time step is computed using (19), where the current energy content of the BESS is normalized by its installed capacity. Equation (20) introduces binary variables ensuring only one SOC band is active at any given time. Equation (21) enforces the logical consistency of SOC band assignment by activating the lower and upper SOC bounds associated with the selected band. Finally, (22) replaces (16), and restricts the charging power based on the active SOC band. It ensures that as the SOC increases, the maximum allowable charging power decreases proportionally, thereby mimicking realistic tapered charging behavior.

## IV. CASE STUDIES

To evaluate the impact of SOC-dependent TCP constraints on long-term microgrid planning, we conducted four case studies using a Pyomo-based optimization framework. The model was solved using GUROBI 11.0.3, with a MIPGAP of 0.0% and a time limit of 3600 seconds. Each case varies in terms of whether tapering constraints are enabled and whether sizing variables (PV and BESS) are fixed or free for the optimizer to choose. These cases are presented in Table II.

TABLE II
SUMMARY OF CASE STUDY SCENARIOS AND OBJECTIVES

| Case No. | Title | Tapering | Sizing | Purpose / Hypothesis |
|---|---|---|---|---|
| 1 | Tapering Enabled | Enabled | Free | Observe oversizing due to tapering; expect higher cost |
| 2 | Tapering Disabled | Disabled | Free | Expected lower objective cost due to no limitation on charging. |
| 3 | Disabled + Isolated | Disabled | Fixed to Case1 sizes | Tests to see if the same sizes perform better without tapering. |
| 4 | Enabled + Constrained | Enabled | Fixed to Case2 sizes | Stress test: Can tapering degrade reliability when not sized for it? |

All four of these cases are tested on a synthetic load demand and PV capacity factor profile, designed specifically to capture a fast-charging requirement that can be practically observed in commercial EVs, frequency regulation applications etc. The profile features a midday PV generation peak between hours 11–14, reaching 1 p.u., while the load remains constant at 0.5 MW during this period and before. A sharp increase in demand occurs during evening hours (17–22), peaking at 1.5 MW when PV output is zero, highlighting a typical solar-load mismatch scenario.

The simulation results for this profile are presented in Fig. 3 and summarized in Table III. For all test cases, each day is treated independently, with no SOC carryover across days. For comparison purposes, the initial SOC of the battery was set to 50% of its rated capacity (i.e., $E_{BESS}^{init} = 0.5 * S_{BESS}$).

TABLE III
SUMMARY OF SIMULATION RESULTS FOR CASE STUDY SCENARIOS

|  | Case1: Tapering Enabled | Case2: Tapering Disabled | Case3: Disabled Isolated | Case4: Enabled Constrained |
|---|---|---|---|---|
| Objective cost | $12,816,525 | $12,323,481 | $12,816,525 | N/A |
| PV size | 3.526 MW | 3.986 MW | 3.526 MW | N/A |
| BESS size | 13.75 MW | 11.0 MW | 13.75 MW | N/A |
| Total load | 164,250 MWh | 164,250 MWh | 164,250 MWh | N/A |
| PV generation | 114,353 MWh | 129,293 MWh | 114,353 MWh | N/A |
| PV curtailed | 12,466 MWh | 14,990 MWh | 9045 MWh | N/A |
| BESS charging | 83,637 MWh | 96,053 MWh | 87,059 MWh | N/A |
| BESS discharging | 146,000 MWh | 146,000 MWh | 146,000 MWh | N/A |
| Load Shedding | 0.0 MWh | 0.0 MWh | 0.0 MWh | 14,238 MWh |

Results from the synthetic profile reveal notable differences across these cases. Case 1, with tapering enabled, resulted in an objective cost of $12.82 million and required a BESS size of 13.75 MWh. In contrast, Case 2, without tapering, achieved a lower objective cost of $12.32 million with only 11 MWh of BESS capacity. The increased cost in Case 1 reflects the larger BESS needed to meet charging limitations imposed by tapering, highlighting the trade-off between efficiency and charging flexibility. Case 3 used the same investment decisions as Case 1 (which was optimized with tapering enabled) but ran without tapering constraints. Interestingly, it resulted in the same objective cost as Case 1. This reflects that the current cost structure in the model does not penalize fast charging or efficiency losses, factors that tapering is intended to address. If the model included penalties for aggressive charging (e.g., accelerated degradation or reduced efficiency), the benefits of tapering would become evident. Finally, Case 4 exhibited significant load shedding when tapering was enforced on a system originally sized without it. This clearly demonstrates that neglecting tapering during planning can compromise reliability, especially under tight sizing margins or aggressive charging demands.

## V. CONCLUSION

This work introduced a tractable method to approximate CCCV charging behavior in long-term microgrid planning through SOC-dependent TCP constraints. By dynamically reducing allowable charging power at high SOC levels, the proposed approach captures the trade-off between reduced charging flexibility and efficiency gain. A MATLAB-based proof-of-concept demonstrated the technical validity of tapering, showing that energy losses become more significant as internal resistance increases, a trend especially relevant for second-life batteries. Integrating these tapering constraints into a long-term optimization model revealed their impact on BESS sizing, cost, and reliability under various scenarios.

Simulation results showed that under a dynamic load and solar profile, tapering leads to noticeable oversizing of the BESS and increased objective cost, highlighting its importance in planning models exposed to fast-charging demands. Importantly, ignoring tapering during planning and enforcing it during operation can result in significant load shedding, underscoring the reliability risks of omitting this behavior.

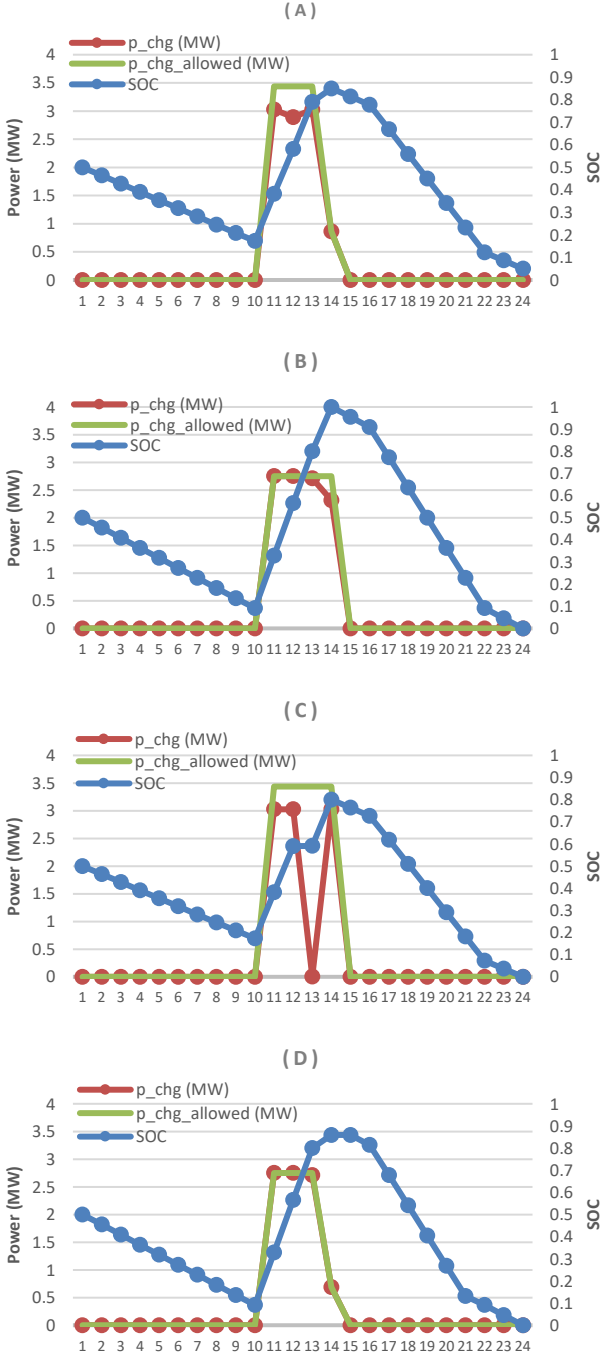

Fig. 3: Charging power and SOC profiles under case study scenarios (A) Tapering Enabled, (B) Tapering Disabled, (C) Disabled + Isolated, and (D) Enabled + Constrained. The plots show actual charging power, tapering-imposed power limits, and resulting SOC profiles across the day.

Future work will focus on refining the model by linearizing the current mixed-integer quadratically constrained into a mixed-integer linear programming problem to enhance computational performance and scalability. Additional research is also planned to validate the proposed framework in a high-demand practical use cases such as frequency regulation and EV fleet charging, where fast charging plays a critical role. Incorporating C-rate-dependent efficiency and degradation models will further improve realism, especially when assessing second-life battery applications. Finally, increasing the time resolution from hourly to sub-hourly (e.g., 15-minute intervals) may enable more accurate modeling of tapering dynamics and operational flexibility.